\documentclass{elsarticle}
\usepackage[margin=3.4cm]{geometry}
\usepackage{latexsym}
\usepackage{longtable}
\usepackage{float}
\usepackage[latin2]{inputenc}
\usepackage{graphicx}
\usepackage{ae}
\usepackage{aecompl}
\usepackage{setspace}
\usepackage{amsmath}
\usepackage{amssymb}
\usepackage{amsthm}
\usepackage[hidelinks]{hyperref}

\makeatletter
\def\ps@pprintTitle{%
	\let\@oddhead\@empty
	\let\@evenhead\@empty
	\def\@oddfoot{\centerline{\thepage}}%
	\let\@evenfoot\@oddfoot}
\makeatother

\begin{document}

\title{The Frequent Complete Subgraphs in the Human Connectome}
	
\author[p]{Máté Fellner}
\ead{fellner@pitgroup.org}
\author[p]{Bálint Varga}
\ead{balorkany@pitgroup.org}
\author[p,u]{Vince Grolmusz\corref{cor1}}
\ead{grolmusz@pitgroup.org}
\cortext[cor1]{Corresponding author}
\address[p]{PIT Bioinformatics Group, Eötvös University, H-1117 Budapest, Hungary}
\address[u]{Uratim Ltd., H-1118 Budapest, Hungary}

\date{}

\begin{abstract}
	While it is still not possible to describe the neural-level connections of the human brain, we can map the human connectome with several hundred vertices, by the application of diffusion-MRI based techniques. In these graphs, the nodes correspond to anatomically identified gray matter areas of the brain, while the edges correspond to the axonal fibers, connecting these areas. In our previous contributions, we have described numerous graph-theoretical phenomena of the human connectomes. Here we map the frequent complete subgraphs of the human brain networks: in these subgraphs, every pair of vertices is connected by an edge. We also examine sex differences in the results. The mapping of the frequent subgraphs gives robust substructures in the graph: if a subgraph is present in the 80\% of the graphs, then, most probably, it could not be an artifact of the measurement or the data processing workflow. We list here the frequent complete subgraphs of the human braingraphs of 414 subjects, each with 463 nodes, with a frequency threshold of 80\%, and identify 812 complete subgraphs, which are more frequent in male and 224 complete subgraphs, which are more frequent in female connectomes. 
\end{abstract}

\maketitle

\section*{Introduction}

Diffusion MRI-based macroscopic mapping of the connections of the human brain is a technology that was developed in the last 15 years \cite{Hagmann2012,Seung2009,Sporns2005,Lichtman2008a}. Applying the method, we are able to construct braingraphs, or connectomes, from the diffusion MRI images \cite{Hagmann2012, Daducci2012, Kerepesi2016b}: the vertices of the graph are anatomically labeled areas of the gray matter (called ``Regions of Interests'', ROIs), and two such ROIs are connected by an edge, if a complex workflow, involving either deterministic or probabilistic tractography, finds axonal fibers between the two ROIs. Therefore, one can construct graphs, with up to 1015 nodes and several thousand edges, from the MR image of each subject. 

The analysis of these graphs is a fast-developing and an important area today: these connections form the ``hardware'' of all brain functions on a macroscopic level \cite{Seung2009,Sporns2005,Lichtman2008a}. Naturally, it would be exciting to map the neuronal scale human connectome, too: here the nodes were the individual neurons, and two nodes (or neurons), say X and Y, would be connected by a directed edge, say $(X,Y)$,  if the axon of $X$ were connected to a dendrite of $Y$. Unfortunately, to date, the neural-level connectome of only one adult organism is described: that of the nematode {\em Caenorhabditis elegans} with 302 neurons, in the year 1986 \cite{White1986}. In larval state, two more neuronal level connectomes are published: the larva of the fruitfly {\em Drosophila  melanogaster} \cite{Ohyama2015}, and the tadpole larva of the {\em Ciona intestinalis} \cite{Ryan2016}. Despite of some exciting, very recent developments \cite{Zheng2018}, the complete connectome of the adult {\em Drosophila  melanogaster} with 100,000 neurons, is not determined yet. Humans have 80 billion neurons in their brains. Therefore, the mapping and the analysis of the neuronal scale human connectome is out of our reach today. 

There are numerous results  published for the analysis of the diffusion MRI-computed human connectomes, e.g., \cite{Hagmann2012,Seung2009,Sporns2005,Lichtman2008a}. Our research group has also contributed some more graph-theoretically oriented analytical methods, like the comparison of the deep graph theoretical parameters of male and female connectomes \cite{Szalkai2015,Szalkai2016a,Szalkai2015c}, the parameterizable human consensus connectome \cite{Szalkai2015a,Szalkai2016}, the description of the individual variability in the connections of the major lobes \cite{Kerepesi2015c}, the discovery of the Consensus Connectome Dynamics \cite{Kerepesi2015b,Kerepesi2016,Szalkai2016d,Szalkai2016e} the description of the frequent subgraphs of the human brain \cite{Fellner2017}, and the Frequent Neighborhood Mapping of the human hippocampus \cite{Fellner2018}. 

The reasons of our graph-theoretical approach are listed as follows:

\begin{itemize}
	\item[(i)] Graph theory has a long history of great successes, starting with the paper of Leonhard Euler in 1741 on Königsberg's bridges \cite{Eulera}.
	\item[(ii)] ``Pure mathematical'' graph theory and its applications in computer engineering reached an exceptionally high level of development in the late XX$^{th}$ and in the early XXI$^{st}$ century, just mentioning three famous examples: the Strong Perfect Graph Theorem \cite{Chudnovsky2002}, Szemerédi's Regularity Lemma \cite{Szemeredi1975}, and the intricate parallel algorithms for multiprocessor routing in \cite{Leighton2014}.
	\item[(iii)] Graph theoretical definitions and notations are well-developed, clear, and usually catch the deep and most relevant properties of the networks examined.
\end{itemize}

\subsection*{Frequent edges and subgraphs: a robust analysis} The data acquisition and processing workflow, whose results are the braingraphs or structural connectomes, has numerous delicate steps. Naturally, errors may occur in MRI recordings and processing, as well as in segmentation, parcellation, tractography and graph computation steps \cite{Hagmann2012,Jbabdi2011,Mangin2013}. When we have hundreds of high-quality MR images, we can analyze the {\em frequently appearing} graph edges or subgraphs, in order to derive robust, reproducible results, appearing in high fraction of the brains imaged. By analyzing only the frequently appearing structural elements, the great majority of data acquisition and processing errors will be filtered out.

Our first effort for describing frequent edges in human connectome was the construction of the Budapest Reference Connectome Server \cite{Szalkai2015a,Szalkai2016}, in which the user can select the frequency threshold $k\%$ of the edges, and the resulting consensus connectome contains only those edges, which are present in at least $k\%$ of the subjects. The generated consensus connectome can be both visualized and downloaded at the site \url{https://pitgroup.org/connectome/}. 

The frequent, connected subgraphs of at most 6 edges are mapped in the human connectome in \cite{Fellner2017}. The frequencies were compared between female and male connectomes, and strong sex differences were identified: there are connected subgraphs, which are significantly more frequent in males than in females, and there are a higher number of connected graphs that are more frequent in females than in males. 

The direct connections of important brain areas are of special interest: correlations between the present or missing connections and psychological tests results or biological parameters may enlighten the fine structure-function relations of our brain. For error-correction reasons, the frequent neighbors of the relevant brain areas form the robust objects of study: small errors in the data processing workflow will most probably have no effects on the frequent connections. In our work \cite{Fellner2018}, we have introduced the method of the Frequent Neighborhood Mapping, which describes the frequent neighbor sets of the given nodes of the braingraph. In \cite{Fellner2018}, we have demonstrated the method by mapping the frequent neighborhoods of the human hippocampus: one of the most deeply studied part of the brain. We have mapped the frequent neighbor sets of the hippocampus, and we have found sex differences in the frequent neighbor sets: males have much more frequent neighbor sets of the hippocampus than the females; therefore, the neighborhoods of the men's hippocampi are more regular, with less variability than those of women. This observation is in line with the results of \cite{Szalkai2015,Szalkai2016a,Szalkai2015c}, where we have shown that the female connectomes are better expander graphs than the braingraphs of men.

In the present contribution, we are mapping the frequent complete graphs of the human connectome, based on the large dataset of the Human Connectome Project \cite{McNab2013}. Our dataset contains the braingraphs of 414 subjects. A recently appeared work \cite{Sizemore2018} deals with complete subgraphs in braingraphs of 8 subjects, each with 83 nodes. Our results are derived from 414 braingraphs, each of 463 nodes. Therefore, we are able to find frequent structures, i.e., frequent complete subgraphs in our dataset of 414 graphs (while it is not feasible to derive frequent structures from 8 graphs).

\subsection*{Cliques vs. complete subgraphs} 
Here we intend to clarify some graph theoretical terms. A complete graph on $v$ vertices contains (undirected) edges, connecting all the ${v\choose2}=v(v-1)/2$  vertex-pairs: that is, in a complete graph, each pair of vertices are connected by an edge. 

If we have a graph $G$ on $n$ vertices, we can look for the complete subgraphs $H$ of $G$: all the vertices and the edges of $H$ need to be vertices and edges of $G$ (i.e., $H$ is a subgraph of $G$), and, moreover, $H$ needs to be a complete graph. 

The complete subgraph of the maximum vertex-number of $G$ is called a clique. The clique number of graph $G$, $\omega(G)$, is the number of vertices  in the largest complete subgraph of $G$. Computing the clique number $\omega(G)$ is a well-known hard problem: it is NP-hard \cite{GJ1979}, that is, it is not probable that one could find a fast (i.e., polynomial-time) algorithm for computing  $\omega(G)$. Moreover, in general, not only the exact value of $\omega(G)$ is hard to compute, but it is also very difficult to approximate, even roughly \cite{Hastad1996}. In special cases, however, when the number of the vertices is only several hundred, and the graph is not too dense, that is, it has not too many edges, then all the frequently appearing complete subgraphs can be computed relatively quickly by the {\em apriori} algorithm \cite{Assoc,Agrawal1994}. The computational details are given in the Materials and Methods section.

Our goal in the present contribution is to map the frequently appearing complete subgraphs in human connectomes. We need to make clear that our analysis is done on 463-vertex braingraphs. Therefore, if a complete subgraph is found, it does not imply the neuronal level existence of complete subgraphs. It implies, however, that the macroscopic ROIs, corresponding to the vertices of the complete graphs discovered, are connected densely to each another, probably even on the neuronal level. 

In the literature one may find numerous references to the ``rich club property'' of some networks, related to the braingraph \cite{Ball2014,vandenHeuvel2011}. Here we prefer using classical graph theoretical terms instead of this ``rich club property'', consequently, we intend to map those densely connected subgraphs of the human connectomes, which form complete graphs, and appear in at least the 80\% of the all braingraphs considered.

\section*{Discussion and Results}

First we review the frequent complete subgraphs of the human braingraph, next we analyze the significant differences in their frequencies in males and in females.

\subsection*{Frequent complete subgraphs of the human connectome}

Supporting Table S1 contains the complete subgraphs of the human connectomes appearing in at least 80\% of the graphs of the 414 subjects examined. In each row, the vertices of the complete subgraphs are listed, together with their frequencies of appearance. Note, that the vertices of a complete graph uniquely determine its edges.  The list is redundant in the following sense: if a $k$-vertex complete graph has frequency at least 80\%, then all of its complete subgraphs are also listed. We find that this redundancy helps in the analysis of the results, as it will be clear from what follows.

We would like to emphasize the following very simple, but powerful fact: If a given subgraph $U$ has a frequency, say $\ell\%$, then all subgraphs of $U$ has frequency at least $\ell \%$. This is the central point in the apriori algorithm \cite{Assoc,Agrawal1994}, and it was noted and applied in \cite{Fellner2017,Fellner2018}.

The ROIs in Table S1 carry the names of the resolution-250 parcellation labels (where the number 250 refers to the approximate number of vertices in each hemisphere; the graphs of resolution-250 contain 465 vertices, not just 250), based on the Lausanne 2008 brain atlas \cite{Hagmann2008} and computed by using FreeSurfer \cite{Fischl2012} and CMTK \cite{Gerhard2011,Daducci2012}, given at 
\url{https://github.com/LTS5/cmp_nipype/blob/master/cmtklib/data/parcellation/lausanne2008/ParcellationLausanne2008.xls}. The ``lh'' and the ``rh'' prefixes abbreviate the ``left-hemisphere'' and ``right-hemisphere'' terms of localizations.

\subsection*{Complete subgraphs appearing in each subject} 
Here we list the maximal complete subgraphs from supporting Table S1, which are present in all of the braingraphs, and contains at least three nodes: 

\bigskip

{\small 
 \noindent L1: (Left-Caudate)(Left-Pallidum)(Left-Putamen)(Left-Thalamus-Proper), 
  
 \noindent L2: (Left-Hippocampus)(Left-Putamen)(Left-Thalamus-Proper), 
  
 \noindent L3: (Left-Putamen)(Left-Thalamus-Proper)(lh.insula\_1)

\noindent  R1: (Right-Caudate)(Right-Pallidum)(Right-Putamen)(Right-Thalamus-Proper)
  
\noindent  R2: (Right-Hippocampus)(Right-Putamen)(Right-Thalamus-Proper)
  
\noindent  R3: (Right-Putamen)(Right-Thalamus-Proper)(rh.insula\_2)
  
\noindent  R4: (rh.superiorfrontal\_7)(rh.superiorfrontal\_8)(rh.superiorfrontal\_9)}

\medskip

Note that L1 and L2 correspond to R1 and R2, and L3 almost corresponds to R3. Complete graph R4 has no correspondence in the left hemisphere (which are present in each subject), but in the left hemisphere, the superiorfrontal regions are also connected densely, as one can verify easily from Table S1.

We believe that the connections between the above-listed areas are very strong in each subject: so strong that they are not affected by measurement errors and individual variability.

\subsection*{The largest frequent complete subgraphs}

The largest complete subgraphs, which are present in at least the 80\% of the subjects, have seven vertices, and they are located in the left hemisphere. The first one connects the left putamen with six vertices in the left frontal lobe (B1), the second one connects the left caudate and the left putamen ROIs to five left frontal areas (B2):

{\small 
	\noindent B1: (Left-Putamen) (lh.lateralorbitofrontal\_4) (lh.lateralorbitofrontal\_6) (lh.lateralorbitofrontal\_7) (lh.parstriangularis\_3) (lh.rostralmiddlefrontal\_12) (lh.rostralmiddlefrontal\_9) 
	\medskip
	
	\noindent B2:  (Left-Caudate) (Left-Putamen) (lh.lateralorbitofrontal\_7) (lh.medialorbitofrontal\_2) (lh.rostralanteriorcingulate\_1) (lh.rostralmiddlefrontal\_12) (lh.rostralmiddlefrontal\_9)
\medskip 

There are 48 different 6-vertex complete subgraphs, which are present in at least 80\% of the connectomes. Only 6 of these are situated in the right hemisphere, the other 42 are in the left hemisphere. 

\medskip
\subsection*{Complete subgraphs across the hemispheres}

 Since the neural fiber tracts, connecting the two hemispheres of the brain, are very dense in the corpus callosum, their tractography in the diffusion MR images is difficult since the fiber-crossings cannot always be tracked reliably \cite{Reginold2016, Hofer2006}. 
 
 We have found only relatively few frequent complete subgraphs of the human connectome, which have nodes from both hemispheres. Here we list those, which are present in more than 80\% of the braingraphs studied; therefore, they are most probably not false positives. Again, we are listing only the maximal complete subgraphs for clarity. We note that most ROIs in the list are the parts of the striatum: each complete subgraph contains either a caudate nucleus or a nucleus accumbens of either the right- or the left hemisphere:
 
 \bigskip
 
 {\footnotesize
 \noindent A1: (Left-Accumbens-area)(Left-Caudate)(Left-Thalamus-Proper)(Right-Caudate)
 
 \noindent A2: (Left-Accumbens-area)(Left-Caudate)(Right-Caudate)(lh.rostralanteriorcingulate\_1)
 
 \noindent A3: (Left-Accumbens-area)(Left-Thalamus-Proper)(Right-Thalamus-Proper)
 
 \noindent A4: (Left-Caudate)(Left-Thalamus-Proper)(Right-Caudate)(Right-Thalamus-Proper)
 
 \noindent A5: (Left-Caudate)(Right-Caudate)(lh.caudalanteriorcingulate\_1)(lh.caudalanteriorcingulate\_2)
  
 \noindent A6: (Left-Caudate)(Right-Caudate)(lh.caudalanteriorcingulate\_1)(lh.rostralanteriorcingulate\_1)
  
 \noindent A7: (Left-Caudate)(Right-Caudate)(rh.caudalanteriorcingulate\_1)
   
 \noindent A8: (Left-Caudate)(Right-Caudate)(rh.rostralanteriorcingulate\_2)
     
 \noindent A9: (Left-Thalamus-Proper)(Right-Accumbens-area)(Right-Thalamus-Proper)}
    
 \medskip   
      
 \subsection*{Counts of the hippocampus, thalamus, putamen, pallidum and the amygdala in the frequent complete subgraphs}

 In this section we count the appearances of certain ROIs in the frequent complete subgraphs, with a frequency threshold of 80\%. Our results show that there are considerable differences between the hemispheres in these numbers: The right hippocampus and the right amygdala are present in much more complete subgraphs than the left ones; the left thalamus-proper, the left putamen and the left pallidum are present in much more complete subgraphs than the right ones (Table 1).

 \begin{table}
 	\centering
 	\begin{tabular}{ | l | r | r | }
 		\hline
 		
 		& Left & Right  \\ \hline \hline
 		Hippocampus&  187& 247\\  \hline
 		Amygdala & 66& 99\\ \hline
 		Thalamus-Proper & 265& 175\\ \hline
 		Putamen & 1041& 673\\ \hline
 		Pallidum & 149& 123\\ \hline
 	\end{tabular}
 	\caption{The number of appearances of the hippocampus, the amygdala, the thalamus-proper, the putamen and the pallidum in the frequent complete subgraphs, with a frequency threshold of 80\%, in each hemisphere. The right hippocampus and the right amygdala are present in much more complete subgraphs than the left ones; the left thalamus-proper, the left putamen and the left pallidum are present in much more complete subgraphs than the right ones.}
 \end{table}

\subsection*{Sex differences}

Mapping sex differences in the human connectome is a hot and fast-developing area of research. In our earlier works we have shown - first in the literature - that in numerous well-defined graph theoretical parameters, women have ``better connected'' braingraphs than men \cite{Szalkai2015,Szalkai2016a,Szalkai2015c}.  In the work \cite{Fellner2017} we have mapped the frequent subgraphs of the human brain of at most 6 vertices, and have found sex differences: there are numerous frequent connected subgraphs, which are more frequent in men than in women, and, similarly, which are more frequent in men than in women. In the study of \cite{Fellner2018}, we have mapped the neighbor-sets of the human hippocampus and found also significant sex differences in these sets.

Here we compare the frequencies of the complete subgraphs of the connectomes of men and women. We have found significant differences in the frequencies of some complete subgraphs, with the advantage at men and women, too.

We have found much more complete subgraphs with significantly higher frequency in men than in women. More exactly, Supporting Table S2 lists 224 complete subgraphs, with significantly higher frequency in females than in males, while Table S3 lists 812 complete subgraphs, where their frequencies in males were higher than in females (with p=0.01, and the inclusion threshold was a minimum 80\% for the larger frequency).

This observation, in a sense, shows that men's connectomes show less inter-personal variability in complete subgraphs than those of women. This observation is in contrast with our findings in \cite{Fellner2017}, where we have shown that women have much more 6-vertex frequent subgraphs than men: but in \cite{Fellner2017} we required connectedness, and not completeness.

\begin{figure}[H]
	\begin{center}
		\includegraphics[width=12cm]{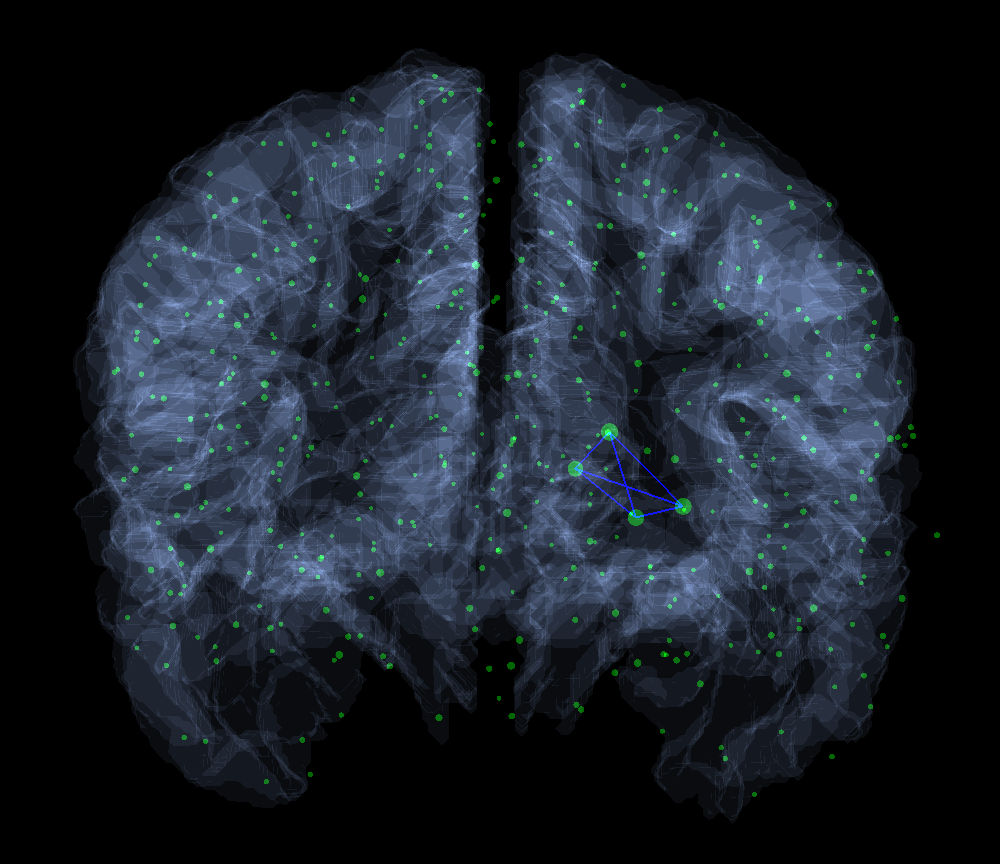}
		\caption{A complete subgraph with four vertices, which is present in all 414 human connectomes examined.  The vertices of the depicted subgraph are Right-Caudate, Right-Pallidum, Right-Putamen, Right-Thalamus-Proper. The supporting Table S1 contains all the complete subgraphs with frequency of at least 80\%, Table S2 contains the complete subgraphs, where the frequency of their appearance in females is significantly higher (p=0.01) than in males; Table S3 contains those, where the frequency is significantly higher in males than in females. The supporting tables are available at \url{http://uratim.com/cliques/tables.zip}. }
	\end{center}
\end{figure}

\section*{Materials and Methods}

\subsection*{The Data Source and the Graph Computation}

The data source of the present study is the website of the Human Connectome Project at the address \url{http://www.humanconnectome.org/documentation/S500} \cite{McNab2013}. The dataset contains the HARDI MRI data of healthy human young adults between the ages of 22 and 35 years.

The CMTK toolkit \cite{Daducci2012}, together with the FreeSurfer tool and the MRtrix tractography processing tool \cite{Tournier2012} were applied in the graph generation. In the MRtrix tool, we have applied randomized seeding and deterministic streamline method, with 1 million streamlines. We have studied here graphs with 463-vertex resolution. The parcellation data is given in the CMTK nypipe GitHub repository \url{https://github.com/LTS5/cmp_nipype/blob/master/cmtklib/data/parcellation/lausanne2008/ParcellationLausanne2008.xls}.

Further particularities of the graph processing workflow are described in 
 \cite{Kerepesi2016b}, where the \url{http://braingraph.org} repository is also given. The braingraphs, analyzed here, can be accessed  
at the \url{https://braingraph.org/cms/download-pit-group-connectomes/} site, choosing the ``Full set, 413 brains, 1 million streamlines'' option. 

\subsection*{The Algorithm} In general, in a graph $G$, finding the size of the largest complete subgraph, called the {\em clique-number}, and denoted by $\omega(G)$, is a hard problem: it is NP-hard \cite{GJ1979}. Naturally, finding the largest complete subgraph itself cannot be easier than finding its size $\omega(G)$; therefore, it is also hard. 

Finding the largest complete subgraphs in sparse graphs (i.e., graphs with relatively few edges, compared to the number of its vertices) is usually not a very difficult task, since in these graphs, regularly, there are not too many large complete subgraphs. Finding only the frequently appearing complete subgraphs further simplifies the computational tasks, and we can apply an algorithm, which resembles in many points to the apriori algorithm for finding frequent item sets \cite{Agrawal1994,Assoc}, and this algorithm is very fast in the practice. 

Now we describe the algorithm: A frequent complete subgraph is characterized by the list of its vertices and the set of its edges. At the beginning, for an (undirected) edge $(v_i,v_j)$, let these two lists be given as $ ( [v_i,v_j], \{(v_i,v_j)\} )$, where $i<j$. 

In general, the vertices of the complete subgraphs are listed in the increasing order of their indices, and the vertices of each edge are listed also in the increasing order of its indices; otherwise, the particular order of the edges is indifferent, since they are elements of an unordered set, and they are stored also as a set.

Now we describe the generating, ``apriori'' step. Let 
 
$$L_1 = ( [v_1, v_2, ..., v_k], \{ (v_1, v_2), (v_1, v_3), ... (v_{k-1}, v_k) \} )$$
and
$$L_2 = ( [u_1, u_2, ..., u_k], \{ (u_1, u_2), (u_1, u_3), ... (u_{k-1}, u_k) \} )$$
be two frequent complete subgraphs of size $k$. If the first $k-1$ vertices of $L_1$ and $L_2$ are the same, and the last ones differ, we will consider generating a new, $k+1$-vertex complete graph, as follows: if 
$v_1 = u_1, v_2 = u2, ... v_{k-1} = u_{k-1}$ and $v_k\ne u_k$, then, by the notation $v_{k+1}=u_k$, we verify the suitable frequency of the complete graph
$L = ( [ v_1, v_2, ... v_k, v_{k+1}], \{(v_1, v_2), (v_1, v_3), ... (v_{k}, v_{k+1})\})$. 

It is easy to see that in the edge list only the last one, $(v_k, v_{k+1})$ is new, all the others are already the edges of the frequent subgraphs $L_1$ or $L_2$. 

In generating $L$ one needs to make sure that the vertices in the vertex-list are ordered by their indices, and that the frequency of $L$ is above the inclusion threshold. 

The apriori generating step is correct, since if $L$ is frequent, then both $L_1$ and $L_2$ were frequent. Additionally, every $k+1$-vertex complete graph is generated only once, since the vertices are in increasing order: $(v_1, v_2,\ldots,v_{k+1})$ can be generated only from  $(v_1, ... v_{k})$ and $(v_1, ... v_{k-1}, v_{k+1})$.

\subsection*{Statistical Notes}

The frequent complete subgraphs were chosen in the following way: 

First, the subjects were partitioned into two disjoint sets, by the parity of their ID number's second digit from the right. Next, in both sets, the complete graphs with the minimum frequency of 80\% were identified, as it was described in the previous section. Only those complete subgraphs were retained, which have had a minimum frequency of 80\% in {\em both} sets under consideration. Then the frequency of these subgraphs were re-calculated for the whole dataset: these frequencies are given in the supplementary tables.

In the computation of sex differences, we have applied $\chi^2$ tests to identify significant differences in the frequencies of the complete subgraphs. Our null hypothesis was that the frequencies are the same in males and females, and we refute this hypothesis with p=0.01. The secondary statistical errors were handled by Holm-Bonferroni corrections \cite{Holm1979}. The un-corrected and the corrected p values are listed in supplementary Tables S2 and S3.

\section*{Conclusions}

By an apriori-like algorithm, we have mapped the frequent (>80\%) complete subgraphs of 414 subjects, each with 463 vertices. The largest frequent complete subgraph has 7 vertices. Most of the largest frequent subgraphs are located in the left hemisphere.  We have also identified the frequent complete subgraphs, containing vertices from both hemispheres, and identified complete subgraphs with significant frequency-differences between the sexes. We have found that men have much more frequent complete subgraphs than women: this result contrasts our earlier finding \cite{Szalkai2015}, where we have shown that women have much better connectivity-related parameters in their connectomes than men in a following sense: while women have better connected braingraphs than men (as it is very precisely described in \cite{Szalkai2015}), the dense subgraphs of men show less inter-individual variability than in women.

\section*{Data availability} The data source of this work is published at the Human Connectome Project's website at \url{http://www.humanconnectome.org/documentation/S500} \cite{McNab2013}. The parcellation data, containing the anatomically labeled ROIs, is listed in the CMTK nypipe GitHub repository \url{https://github.com/LTS5/cmp_nipype/blob/master/cmtklib/data/parcellation/lausanne2008/ParcellationLausanne2008.xls}.

The braingraphs, computed by us, can be accessed at the  \url{https://braingraph.org/cms/download-pit-group-connectomes/} site, by choosing the ``Full set, 413 brains, 1 million streamlines'' option. Here we have used exclusively the 463-node graphs.

The Supplementary Tables are available on-line at the address \url{http://uratim.com/cliques/tables.zip}. Table S1 contains the list of all the complete subgraphs of 414 human connectomes with a minimum frequency of 80\%.  Table S2 contains the complete subgraphs, where the frequency of their appearance in females is significantly higher (p=0.01) than in males; Table S3 contains those, where the frequency is significantly higher in males than in females. In both Tables S2 and S3 a frequency cut-off 80\% is applied to the larger frequency of the appearance in the sexes: only those significant differences are listed, where the larger of the frequencies of males and females are at least 80\%. 

\section*{Acknowledgments}
Data were provided in part by the Human Connectome Project, WU-Minn Consortium (Principal Investigators: David Van Essen and Kamil Ugurbil; 1U54MH091657) funded by the 16 NIH Institutes and Centers that support the NIH Blueprint for Neuroscience Research; and by the McDonnell Center for Systems Neuroscience at Washington University. VG and BV were partially supported by the VEKOP-2.3.2-16-2017-00014 program, supported by the European Union and the State of Hungary, co-financed by the European Regional Development Fund, VG and MF by the NKFI-126472 and NKFI-127909
 grants of the National Research, Development and Innovation Office of Hungary. BV and MF was supported in part by the EFOP-3.6.3-VEKOP-16-2017-00002 grant, supported by the European Union, co-financed by the European Social Fund.
\bigskip 

\noindent Conflict of Interest: The authors declare no conflicts of interest.

\section*{Author Contribution} MF analyzed braingraphs and developed statistical and graph theoretical tools, BV constructed the image processing system and computed the braingraphs, VG has initiated the study, analyzed data and wrote the paper.


\bibliography{v:/vince/CIKKEK/medl}
\bibliographystyle{unsrtnat}

\end{document}